\documentclass[aps,prapplied,twocolumn,superscriptaddress,longbibliography,nofootinbib]{revtex4-2}
\usepackage{amsmath,amssymb,graphicx,bm}
\graphicspath{{./}}
\begin{document}
\title{Blind directions of physical learning networks: where to measure and what to measure}
\author{Quoc-Bao Nguyen}
\affiliation{Department of Bridge and Road Engineering, Hanoi University of Civil Engineering,
55 Giai Phong Street, Bach Mai ward, Hanoi, Viet Nam}
\author{Thai-Son Vu}
\email{sonvt2@huce.edu.vn}
\affiliation{Department of Bridge and Road Engineering, Hanoi University of Civil Engineering,
55 Giai Phong Street, Bach Mai ward, Hanoi, Viet Nam}
\date{\today}
\begin{abstract}
A physical learning network is read at a few accessible nodes. Every task and learning rule that uses only the steady voltages and currents there, at a fixed operating point, acts through the boundary response map. Changes in the kernel of its Jacobian are blind to first order. A walk along one fiber ran to a 600-step cap, one edge then at 15.2 times its start. A decomposition theorem splits the response Jacobian over the hidden components. The blind dimension adds over components whose surviving slots are disjoint. A boundary-to-boundary edge adds one parameter and deletes one slot. Exposing a hidden node changes only its component. For one hidden node the contribution counts the bipartite components of the non-adjacency graph of its neighbors. For a pocket of $h$ hidden nodes a factor-analysis bound caps what outside electrodes can expose. It is attained when every pocket node meets every neighbor and no edge joins two of those neighbors, so past a threshold further electrodes outside such a pocket expose nothing. An electrode inside such a pocket, when its hidden nodes form a clique, is worth $h-1$ directions where one outside is worth none. Read as vector displacements rather than potentials, the same nodes left no deficit beyond counting in all 90 spring networks tested, each a pocket fully joined to five or more accessible nodes, and nothing blind in 88. What limits the reading is the quantity measured as much as the number of contacts. Maximum-weight spanning forests give a proved upper bound on the blind dimension. The matching equality is proved for one hidden node and conjectured beyond. It holds in all 5,343 components whose maximum could be attained and certified, and the forest count matched the certified blind dimension in 1,448 of 1,500 held-out networks, where the maximum must be searched. A self-learning circuit shows the split on hardware.
\end{abstract}
\maketitle

\section{Introduction}

A material that learns adjusts many internal degrees of freedom while a supervisor reads only a few~\cite{stern2021,stern2023,keim2019}. In resistor networks trained by coupled learning~\cite{stern2021,dillavou2022,dillavou2024}, in elastic networks aged into allosteric responders~\cite{pashine2019,rocks2017,rocks2019}, and in shape-changing metamaterials trained in the laboratory~\cite{du2026}, the training signal enters through a set B of accessible nodes and every measurement is taken at the same set. Between the internal parameters and everything that can be read at B stands one object, the boundary response map \(\Lambda_B\), the Dirichlet-to-Neumann map of the network~\cite{cim1998,boyer2016}, known in circuit theory as its Kron reduction~\cite{dorfler2013}. What the supervisor cannot see through \(\Lambda_B\), no training protocol on B can correct to first order.

The inverse problem for \(\Lambda_B\) has a long history. Curtis, Ingerman, and Morrow characterized the circular-planar networks whose conductances are fully determined by \(\Lambda_B\)~\cite{cim1998}. Colin de Verdi\`ere treated planar networks~\cite{cdv1994}, and Colin de Verdi\`ere, Gitler, and Vertigan gave the complete set of local transformations that leave the response matrix of a circular-planar network unchanged, and with it a classification of the critical ones~\cite{cdv1996}. Lam and Pylyavskyy exhibited the non-uniqueness that appears on the cylinder~\cite{lam2012}. The continuum problem behind these is electrical impedance tomography, where uniqueness from the full Dirichlet-to-Neumann map is a theorem in three dimensions and above~\cite{sylvester1987} and, later, in two~\cite{nachman1996}. Boyer, Garzella, and Guevara Vasquez formulated recoverability for arbitrary graphs as a Jacobian-rank test. They proved that the rank is generically constant, and gave the counting bound \(|B|(|B|-1)/2\) together with a numerical procedure and its noise floor~\cite{boyer2016}. The same Schur complement is studied in power systems, where it splits into a sparse boundary part plus a term of rank at most the number of hidden nodes, and where hidden nodes of degree below three are known to be unidentifiable~\cite{yuan2023}. It appears in algebraic statistics as a marginal precision matrix with latent variables~\cite{chandrasekaran2012}. Draper and co-workers extended the Jacobian formulation to matrix-valued problems on mass--spring networks~\cite{draper2020}. In the physical-learning literature, which grew from equilibrium propagation~\cite{scellier2017}, Stern and co-workers related the physical response of a trained network to its cost~\cite{stern2024,stern2025}, and Guzman and co-workers read learned solutions from the stiffest modes~\cite{guzman2025}. Dillavou and co-workers observed a slow drift of trained conductances and modeled it as a bias acting along flat directions of the cost~\cite{dillavou2025}. Where to put a limited number of sensors so that parameters become identifiable is a question with its own literature, treated as convex selection~\cite{joshi2009} and as observability of a network~\cite{liu2013}. A companion study bounds not what a network hides but what it can reach: Maxwell--Betti reciprocity confines every attainable response block to a subspace of codimension \(p(p-1)/2\) when \(p\) degrees of freedom are both driven and read~\cite{vu2026}. That obstruction is on the forward map and survives with no hidden node, where the one treated here is on the inverse map and vanishes when the network is recoverable. Nearby, a rank condition on the incidence structure governs when equilibrium propagation and coupled learning converge on linear circuits, and its failure is non-generic in the target~\cite{mcginnis2026}, while the conserved sector masses of a circuit fix where on the solution manifold a local rule lands~\cite{dangol2026}. How the flat directions of a task relate to the structure of \(\Lambda_B\), and how much of a trained state is left undetermined by the wiring and by the choice of B, have not been worked out; nor has the dimension of that manifold, which is fixed before any rule is chosen.

Unlike the recoverability studies~\cite{cim1998,boyer2016,cdv1996,lam2012,draper2020}, which ask whether all conductances can be recovered from \(\Lambda_B\), this study asks for the dimension and structure of what cannot be, and how that depends on the graph and on where the accessible nodes sit. That dimension is called the blind dimension throughout, an unordered pair of accessible nodes is called a slot, and a connected component of the hidden subgraph is called a pocket when the point is that electrodes surround it. Four theorems and one conjecture follow. The first: the response Jacobian decomposes over the connected components of the hidden subgraph, the blind dimension adds over components whose surviving measurement slots are disjoint, and a boundary-to-boundary edge adds one parameter and deletes one slot. The second: for one hidden node that component's contribution has a closed form in terms of the non-adjacency graph of its neighbors. The third: for a hidden component of any size the directions that electrodes outside it can expose are bounded by the dimension of a factor-analysis variety, computed by Drton, Sturmfels and Sullivant~\cite{dss2007}. The bound is imported and the sharpness argument is standard; what is new is that the electrical parametrization attains the bound under complete attachment, which is what carries Ledermann's 1937 counting rule~\cite{ledermann1937} over into a statement about where to put an electrode. The same argument gives a bound of the same shape for central-force spring networks, where each accessible node carries a displacement vector instead of a potential, with a determinantal term that is loose. The fourth: maximum-weight spanning forests give a lower bound on the rank of the response Jacobian, so a forest count bounds the blind dimension from above; for a component with one hidden node that bound is an equality. The conjecture: the same equality holds for components with two or more hidden nodes, which is open and was tested on 1,500 held-out networks. One further statement is empirical. On identical wiring the elastic map left no deficit beyond counting in all 90 test networks, each a pocket joined to every one of five or more accessible nodes, while the scalar map of the same graph lost 6.87 directions on average. Together the results say that the limit on reading a learning network is set by three things: which nodes are accessible, how they sit relative to the hidden components, and what quantity is read at them. Trajectories from a self-learning resistor network~\cite{dillavou2024} show the split between visible and blind directions on hardware.

\section{Blind directions and finite fibers}

Let G be a connected graph with M edges, conductances \(k\in\mathbb R^M_{>0}\), weighted Laplacian \(L(k)\), and accessible set \(B\subset V\). The boundary response map is the Schur complement \(\Lambda_B(k)=L_{BB}-L_{BI}L_{II}^{-1}L_{IB}\), where \(I=V\setminus B\) is the hidden set. Its Jacobian has the rank-one form~\cite{boyer2016}

\begin{equation}\partial\Lambda_B/\partial k_e=u_eu_e^{\mathsf T}\end{equation}

where \(u_e\in\mathbb R^{|B|}\) collects the potential drops across edge e when each accessible node in turn is held at unit potential and the others at zero, with the hidden nodes harmonic (Appendix~\ref{app:jacobian}). Any task whose inputs and outputs are steady voltages and currents at B, read at a fixed operating point, has a loss \(\mathcal L=f(\Lambda_B)\) for some function \(f\) of the boundary response map alone. Its gradient is \(J_\Lambda^{\mathsf T}\partial f\) and vanishes on \(\ker J_\Lambda\), the kernel of the Jacobian of \(k\mapsto\Lambda_B(k)\). Its Hessian is \(J_\Lambda^{\mathsf T}(\partial^2 f)J_\Lambda+\sum_\alpha(\partial f/\partial\Lambda_\alpha)\nabla^2\Lambda_\alpha\); the first term vanishes on \(\ker J_\Lambda\) and the second does not in general, because \(\Lambda_B\) is a nonlinear function of k. That term does vanish there whenever \(\partial f=0\), that is at an exactly fitted task, and that condition is sufficient rather than necessary. The subspace \(\ker J_\Lambda\) is blind to first order to every measurement taken at B and to every learning rule driven by such measurements, and its dimension is the blind dimension of the network. Protocols that read transients, that drive the network outside its linear range, or that measure at hidden nodes fall outside this setting. For protocols that impose and read voltages the direction k is blind as well, since \(\Lambda_B(ck)=c\,\Lambda_B(k)\); one current measurement removes it, so only \(\ker J_\Lambda\) is protocol independent.

The kernel of a Jacobian describes blindness to first order. Because the rank of \(J_\Lambda\) is constant on a dense open set of conductances~\cite{boyer2016} (25 on all 200 random draws for the network of Fig. 1), the constant-rank theorem makes the fiber of \(\Lambda_B\) through any point of that set a smooth manifold of dimension \(\dim\ker J_\Lambda\) in a neighborhood of the point. Nothing is claimed about the fiber away from such points. A fiber was followed by a predictor--corrector scheme (Appendix~\ref{app:levelsets}) on the \(4\times4\) periodic lattice of Fig. 1 with eight accessible nodes, along each of the seven kernel directions and both signs. The shortest walk ended after 23 steps when a conductance reached zero. Two walks were stopped at the 600-step cap with all conductances positive; in one of those an edge reached 15.2 times its starting value while \(\Lambda_B\) stayed fixed to a relative \(3.4\times10^{-16}\) (Fig. 1b). Two networks that differ by a factor of fifteen on an edge therefore agree at B to the precision of the continuation, and a manifold of internal states does the same.

The blind dimension depends on where the accessible nodes are, and it falls as more of them are added. On the 16-node lattice with eight accessible nodes, the counting bound~\cite{boyer2016} allows a rank of 28 and a blind dimension of 4. Over all 12,870 eight-node sets the median is 4 and the maximum 10; 74.6\% reach the bound and 25.4\% fall short, by a median of 3 (Fig. 1c). The values taken are 4, 5, 7, 8, 9 and 10; no set has blind dimension 6. The set used in Fig. 1b, and again in the section on electrode placement below, is a deficient one, with rank 25 and blind dimension 7; the generic value is reported alongside throughout.

\begin{figure*}[t]
\centering
\includegraphics[width=\textwidth]{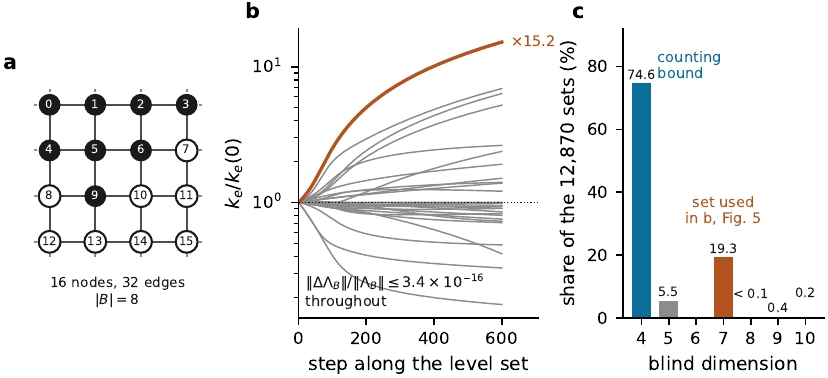}
\caption{\label{fig1}\textbf{a}, The $4\times4$ periodic lattice used throughout, with accessible nodes B (filled) and hidden nodes (open), each carrying the index used in Fig. 5b; dotted stubs are the periodic edges that close the torus. Every measurement at B is a function of the boundary response map $\Lambda_B$, whose Jacobian has the rank-one form of Eq. (1). \textbf{b}, Following one fiber of $\Lambda_B$ on the $4\times4$ periodic lattice with eight accessible nodes: conductance of each edge relative to its starting value along 600 predictor–corrector steps in one kernel direction of the seven; one edge reaches 15.2 times its starting value while $\Lambda_B$ stays fixed to a relative $3.4\times10^{-16}$. \textbf{c}, Distribution of the blind dimension over all 12,870 eight-node accessible sets of the same lattice; the counting bound gives 4, reached by 74.6\% of the sets; the set used in b, and again in Fig. 5, has blind dimension 7.}
\end{figure*}

\section{A decomposition theorem}

The off-diagonal entries of \(\Lambda_B\) determine it, because its rows sum to zero. Call an unordered pair \(\{a,b\}\subset B\) a slot, and call a slot surviving when no boundary-to-boundary edge occupies it. The hidden subgraph \(G[I]\) splits into connected components \(C_1,\dots,C_p\); write \(N(C)\) for the boundary neighbors of C, \(E_C\) for the edges with at least one end in C, and \(E_{\mathrm{int}}(C)\) for the edges with both ends in C. Since \(L_{II}\) is block diagonal over the components (Fig. 2a),

\begin{equation}\Lambda_{ab}(k)=-k_{ab}\,[ab\in E_{BB}]-\sum_C S^C_{ab}(k_{E_C})\quad\text{for } a \ne b\end{equation}

where \(S^C=L_{N(C)C}L_{CC}^{-1}L_{CN(C)}\) depends only on the conductances of \(E_C\) and occupies only slots inside \(N(C)\), and \(E_{BB}\) is the set of boundary-to-boundary edges. Which slots a hidden component can occupy is the edge-creation rule of Kron reduction~\cite{dorfler2013}; what follows is about the rank. Equation (2) has three consequences, proved in Appendix~\ref{app:decomp}.

First, the columns of \(J_\Lambda\) belonging to \(E_{BB}\) are coordinate vectors of slot space. Each such column is a pivot, so a boundary-to-boundary edge adds one parameter and removes its own slot from what the remaining columns are read against. In symbols, \(\operatorname{rank}\,J_\Lambda\) = \textbar{}\(E_{BB}\)\textbar{} + rank \(\Pi[J_{C_1}\cdots J_{C_p}]\), where \(\Pi\) is the projection that deletes the slots occupied by \(E_{BB}\) and \(J_C\) is the Jacobian of the local reduction of component C. The pivot is a statement about the rank. It does not make \(k_{ab}\) identifiable: with \(B=\{a,b\}\), one hidden node \(x\), and edges \(ab\), \(ax\) and \(xb\), the single slot carries \(k_{ab}+k_{ax}k_{xb}/(k_{ax}+k_{xb})\), the rank is 1, and the two-dimensional fiber moves all three conductances. Write \(\delta(C)=m_C-\operatorname{rank}\,\Pi J_C\) for the deficiency of a single component, with \(m_C=|E_C|\) the conductances that touch it. Second, suppose no two hidden components share a surviving slot. Then \(\Pi[J_{C_1}\cdots J_{C_p}]\) is block diagonal, the blind dimension is the sum of the component deficiencies, \(\dim\ker J_\Lambda=\sum_C\delta(C)\), and each \(\delta(C)\) depends only on the subgraph induced by \(C\cup N(C)\) with the boundary edges inside \(N(C)\). Without that hypothesis the sum is only a lower bound. The support of a component is the set of surviving slots it occupies. Of the 4,500 exploration networks, 2,661 had pairwise disjoint supports and the sum rule held in every one. Third, exposing a hidden node changes only the block containing it, and when the surviving slot sets of all components are pairwise disjoint the gains of nodes in different blocks add: checked on 1,192 single-node exposures and on all 631 slot-disjoint pairs in a 300-network sample, without exception (Fig. 2c).

Two corollaries follow. Adding a boundary-to-boundary edge never reduces the blind dimension: it adds one parameter and deletes one slot. And when the surviving slot sets of the components are disjoint, a new electrode can be evaluated on its own hidden component alone; that disjointness is the condition under which the electrode-placement rule below is local. Where two components share a surviving slot the structural change is still confined to one block, but the gain in the global rank is not determined by that block alone.

\section{One hidden node}

Take a hidden component \(C=\{x\}\) consisting of a single node with \(d\) boundary neighbors. Its local reduction is the star--mesh transformation \(S^C_{ab}=k_ak_b/\sum_c k_c\). Let F be the graph on \(N(x)\) whose edges are the pairs not joined by a boundary edge, and let \(b(F)\) be the number of connected components of F that are bipartite, isolated vertices counted as components. Then \(\delta(C)\) = \(b(F)\). The proof (Appendix~\ref{app:onenode}) writes the log-Jacobian of the star--mesh map as \(V-\mathbf 1p^{\mathsf T}\), with V the unsigned incidence matrix of F and \(p\) the vector of attachment conductances normalized to sum to one. The two have the same rank because \(V-\mathbf 1p^{\mathsf T}=V(I-zp^{\mathsf T})\) with \(z=\mathbf 1/2\), and \(\det(I-zp^{\mathsf T})=1-p^{\mathsf T}z=1/2\ne0\), and the rank of an unsigned incidence matrix is \(d-b(F)\)~\cite{grossman1995}.

The formula recovers the classical reductions and goes beyond them (Fig. 2b). A pendant node (\(d=1\)) costs one direction, and so does a series node (\(d=2\)) whose two neighbors are not adjacent; if they are adjacent it costs two. That hidden nodes of degree below three are unidentifiable is known~\cite{yuan2023}; what the formula adds is every remaining case. A degree-3 node with pairwise non-adjacent neighbors costs nothing, the star--triangle transformation, and a node whose neighbors form a clique costs \(d\). A degree-4 node whose neighbors are joined by a perfect matching costs one direction although the counting bound allows zero: the four surviving slots satisfy \(S_{ac}S_{bd}=S_{ad}S_{bc}\), which is the classical condition for a mesh on four terminals to be realizable as a star. The formula was checked on 3,090 single-node components of the exploration ensemble and 400 random graphs with one hidden node, without exception. The forest tests below use a subset of the same ensemble, restricted to components carrying between two and twelve conductances, which holds 2,812 of these single-node components.

\begin{figure*}[t]
\centering
\includegraphics[width=\textwidth]{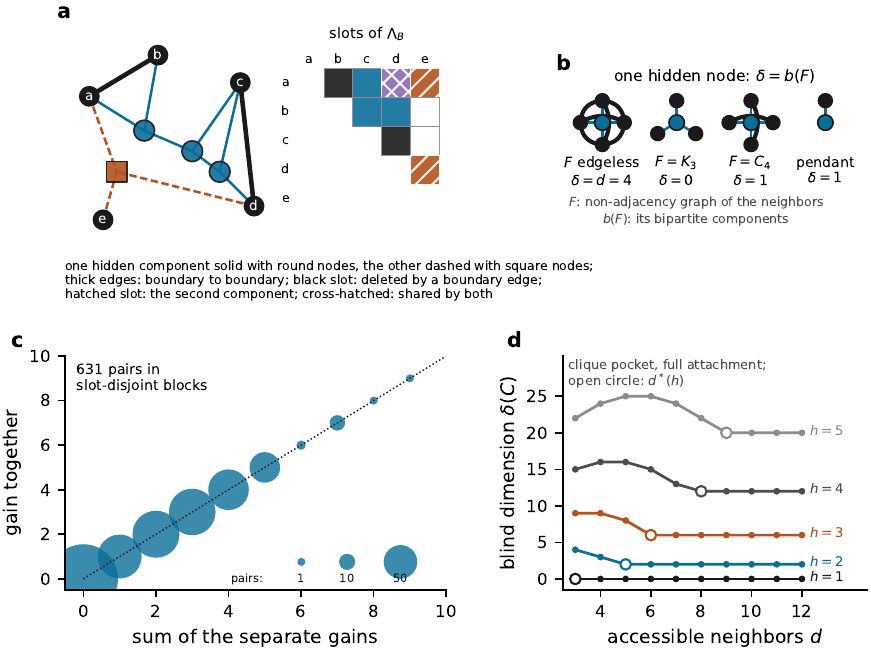}
\caption{\label{fig2}\textbf{a}, Hidden components (shaded) of the hidden subgraph and the slots of $\Lambda_B$ they occupy; boundary-to-boundary edges (thick) delete their slots. \textbf{b}, One hidden node: the component's contribution to the blind dimension equals the number of bipartite components $b(F)$ of the non-adjacency graph F of its neighbors. Four cases: F edgeless (clique of neighbors, $\delta=d$), F a triangle ($\delta=0$, star–triangle), F a 4-cycle (perfect matching of boundary edges, $\delta=1$ although counting allows 0), and a pendant node ($\delta=1$). Checked on 3,090 single-node components. \textbf{c}, Gains from exposing hidden nodes are local and additive. The gain of a node is the number of blind directions its exposure removes. For all 631 pairs of nodes in slot-disjoint components found in a 300-network sample, the gain of exposing both equals the sum of the two separate gains; marker area is the number of pairs at that point. \textbf{d}, Saturation of the blind dimension of a hidden pocket. Each curve is a clique of $h$ hidden nodes joined to every one of $d$ accessible nodes; the open circle marks $d^*(h)=\lceil (2h+1+\sqrt{8h+1})/2\rceil$. Beyond that point every further accessible node adds $h$ conductances and exactly $h$ identifiable directions, so $\delta(C)$ stays at $|E_{\mathrm{int}}(C)|+h(h-1)/2$, which is $h(h-1)$ for a clique. Values are $\delta(C)=m_C-\operatorname{rank}\,\Pi J_C$ computed from the Jacobian, not from the bound.}
\end{figure*}

\section{How much a hidden pocket reveals to electrodes around it}

For a component with more than one hidden node the star--mesh argument no longer applies, but the algebraic shape of the local reduction still constrains the answer. The matrix \(S^C=L_{N(C)C}L_{CC}^{-1}L_{CN(C)}\) is positive semidefinite of rank at most \(h=|C|\), so its off-diagonal part, which is all that reaches slot space, lies in the projection of the symmetric rank-\(h\) locus that drops the diagonal. That projection is the image of the factor-analysis variety under the map that drops the diagonal, and its dimension follows from the dimension of that variety, which is known~\cite{dss2007}. Writing \(d=|N(C)|\) and

\begin{equation}\varphi(h,d)=\begin{cases}\min\big\{hd-\tfrac{h(h-1)}2,\ \tfrac{d(d-1)}2\big\}, & h<d,\\[2pt]\tfrac{d(d-1)}2, & h\ge d,\end{cases}\end{equation}

the local rank obeys the exposure bound \(\operatorname{rank}\,\Pi J_C\le\min\{m_C,\varphi(h,d)\}\). More generally, for any subset \(N_0\subseteq N(C)\) the same holds with \(\varphi(h,|N_0|)\) in place of \(\varphi(h,d)\) plus the number of surviving slots not contained in \(N_0\), and the minimum over \(N_0\) is the bound used below (Appendix~\ref{app:exposure}). Under partial attachment this minimum is only a bound: in a separate family of 960 networks that sweeps the attachment fraction (Appendix~\ref{app:ensembles}), it was attained in 415 of the 564 partially attached cases, while the 396 completely attached cases are all tight by the theorem below. The answer under partial attachment is a generic symmetric matrix completion problem whose sharp theory~\cite{bernstein2021} is not used here. The bound is attained whenever every node of \(C\) is joined to every node of \(N(C)\) and no boundary edge lies inside \(N(C)\): the local rank is then exactly \(\varphi(h,d)\). Both are needed. Dropping the second: with \(h=1\), \(d=4\) and a perfect matching of boundary edges on \(N(C)\) the local rank is 3, not \(\varphi(1,4)=4\). Dropping the first: with \(C=\{x,y\}\) joined by an edge, \(N(C)\) of size four, no boundary edge, and \(x\) joined to three of the four while \(y\) is joined to one, the local rank is 4 where \(\min\{m_C,\varphi(2,4)\}=5\). Sharpness is proved by letting the internal conductances go to zero, where the reduction becomes a sum of \(h\) rank-one terms whose image contains a relatively open piece of the smooth locus of the rank-\(h\) matrices. The bound held in all 711 random components tested and in all 400 held-out networks, and equality under complete attachment held in all 192 test networks with \(h=1\) to \(6\) and \(d=2\) to \(9\).

The design consequence is a saturation law. Take a pocket joined to every accessible node, with no boundary edge inside \(N(C)\). Once \(d\) reaches \(d^*(h)=\lceil(2h+1+\sqrt{8h+1})/2\rceil\), the blind dimension of the component is

\begin{equation}\delta(C)=|E_{\mathrm{int}}(C)|+\tfrac{h(h-1)}2\end{equation}

with \(E_{\mathrm{int}}(C)\) the internal edges of the pocket, and it does not move as further accessible nodes are added: each new one contributes exactly \(h\) conductances and exactly \(h\) identifiable directions (Fig. 2d). The threshold is Ledermann's bound~\cite{ledermann1937}: \(d^*(h)\) is the smallest number of observed variables at which the parameter count of an \(h\)-factor model no longer exceeds the number of observed covariances, which is where \(h\) factors become generically identifiable in the factor-analysis reading~\cite{dss2007}. What is read backwards here is the consequence, since the same threshold is where an electrode outside the pocket exposes nothing. The thresholds are 3 accessible neighbors for one hidden node, 5 for two, 6 for three, 8 for four and 9 for five. For \(h=1\) equation (4) gives zero, which is the one-node formula again, since the non-adjacency graph of a fully attached single node is a complete graph and carries no bipartite component. For a clique of \(h\) hidden nodes it gives \(h(h-1)\). Equation (4) matched the computed rank in all 136 test networks in the saturated regime, across four internal wirings of the pocket: no internal edge, a path, a cycle and a clique, and the onset was at \(d^*(h)\) in each of them. That ensemble runs to \(h=5\); the saturated regime for h = 6 begins at \(d^*(6)=10\) and is covered only by the fully unexposed cases of the interior-electrode family in the next paragraph.

An electrode inside the pocket does expose directions. For a clique of \(h\) hidden nodes joined to every one of \(d\ge d^*(h)\) accessible nodes, with no boundary edge inside \(N(C)\), exposing \(t\) of them leaves \(\delta=(h-t)(h-1)\). This is a corollary of the bound and its sharpness argument, proved in Appendix~\ref{app:exposure}. Exposing a node deletes every slot that meets it, and the parameter count of the smaller pocket then gives \(h-1\) blind directions removed by each interior electrode and none by an exterior one. As a check, the formula agreed with the rank in all 50 cases tested, \(h=2\) to \(6\) and \(d=10\) or \(12\). Forty of the fifty ranks were computed; the ten with every pocket node exposed are full by construction and were not recomputed. The two together are a rule for the next electrode: compare each hidden component's neighborhood size with \(d^*(h)\), and a component past that threshold can be improved only from inside.

On the random ensembles of this paper the new bound is almost never the binding one: hidden components there have few boundary neighbors, so the parameter count or the slot count is the smaller number in 399 of the 400 held-out networks tested. In the one exception the factor-analysis term lowers the bound from 33 to 32, against a true rank of 30. It is the smaller number where electrode design asks for many accessible nodes around a small hidden pocket, where the slot count alone would suggest everything is identifiable.

\section{What a vector reading at the same nodes recovers}

What is read at an accessible node matters as much as which node it is. Reading a displacement vector instead of a potential gives \(D\) numbers per node instead of one, and a \(D\times D\) block of \(D^2\) entries per pair of nodes instead of a single entry, while an added bond still carries one unknown. A bound of the same shape survives the change, though its determinantal term is no longer sharp. The saturation law it produces in the scalar case does not survive. The argument uses only that measurements at accessible nodes are functions of a boundary response map that is a sum of rank-one terms in the parameters. Central-force spring networks share it. The full stiffness matrix is \(\sum_e k_eq_eq_e^{\mathsf T}\) with \(q_e\) the bond-direction vector, and the boundary stiffness map is its Schur complement onto the displacements of the accessible nodes, exactly as \(\Lambda_B\) is the Schur complement of the Laplacian. Its Jacobian therefore has the same rank-one form as Eq. (1), with \(y_e\) the bond-direction-projected drop of the harmonic extension in place of \(u_e\)~\cite{draper2020}. The counting changes, because the boundary map now has \(D(D+1)/2\) rigid zero modes.

Each node carries \(D\) degrees of freedom, so the local reduction of a hidden pocket is a symmetric positive semidefinite matrix of size \(Dd\) with rank at most \(Dh\), and the same determinantal argument gives \(\operatorname{rank}\,\Pi J_C\le\min\{m_C,\ \psi(Dh,Dd),\ (Dd-D_r)(Dd-D_r+1)/2\}\) with \(\psi(r,q)=rq-r(r-1)/2\) for \(r<q\) and \(q(q+1)/2\) otherwise, and \(D_r=D(D+1)/2\) the number of rigid motions. Two of the three terms bind in the tests: the parameter count in 99 of the 120 configurations, the rigid-motion cap in 19, with two ties. The term \(\psi\) binds in none of them, and it never can: as soon as the pocket is rigid when its neighbors are pinned, which forces \(Dd\ge D_r\), the term \(\psi\) is at least the smaller of the other two, at every \(D\) (Appendix~\ref{app:exposure}). The elastic bound is therefore \(\min\{m_C,\ (Dd-D_r)(Dd-D_r+1)/2\}\) in effect, and \(\psi\) marks only where a sharper determinantal term would have to sit. This \(\psi\) is the dimension of the rank-\(Dh\) locus itself, not of its off-block-diagonal projection, so unlike \(\varphi\) it is loose; the block analogue of the factor-analysis count is open. The component's contribution again annihilates rigid motions. Its \(D\times D\) diagonal blocks are therefore determined by the off-diagonal ones, exactly as the scalar diagonal is determined by the zero row sums, and the same block-diagonal projection applies. The third term counts the symmetric matrices on the accessible displacements that annihilate the \(D_r\) rigid motions. The bound assumes the pocket is rigid when its boundary neighbors are pinned. That is a hypothesis here and not, as in the scalar case, automatic: a hidden node of degree one, or one collinear with its two neighbors, violates it. The bound held in all 120 test networks of a family built to order for it (Appendix~\ref{app:springs}), of which the 90 with five or more accessible neighbors carry the comparison below. What changes is which term binds. In the scalar case the determinantal term \(\varphi(h,d)\) falls below the parameter count once the pocket has enough neighbors, and that is where the saturation law comes from. The elastic determinantal term does not fall below it at \(D=2\), because an attachment spring adds one parameter and four observable entries. On the same wiring with \(d\ge5\), the elastic blind dimension equaled the pure counting value \(\max\{0,\ m_C-(2d-3)(2d-2)/2\}\) in all 90 networks tested and was zero in 88, while the scalar blind dimension of the identical graph averaged 6.87 and was strictly larger in 72 of the 90 (Fig. 3a). That is an observation on these 90 networks, not a theorem: only the bound is proved. Below five accessible neighbors the elastic map is deficient too, and the counting value held in 15 of 30.

Twenty-seven two-dimensional networks were generated from a triangular lattice with jittered positions, at mean coordinations between 4.0 and 4.85, with 23 to 40 nodes and 46 to 97 springs (Appendix~\ref{app:springs}). Each was swept twice, growing the accessible set one node at a time outermost first and in random order, giving 54 sweeps; the two sweeps of a graph are not independent, and counts below are over sweeps. A sweep stops once both maps have reached full rank, so the denominator falls with the number of accessible nodes: 54 sweeps reach twelve, 49 reach sixteen and 41 reach twenty. Every sweep that stopped early had reached completeness in both maps, so the sweeps dropped from the denominator are ties and not censored in favor of either map.

On the spring ensemble just described the advantage of the vector reading is largest when the accessible set is small. At twelve accessible nodes the elastic blind dimension is smaller in 43 of 54 sweeps, medians 12.5 against 20.5 (Fig. 3b). The advantage decays as nodes are added: 25 of 49 sweeps at sixteen nodes, medians 10 and 15, then 14 of 41 at twenty, medians 8 and 9. Completeness arrives at nearly the same place in the two maps. With outermost-first ordering the two maps required the same number of accessible nodes in 18 of the 25 networks that reached completeness in both, medians 25 each.

Rank plateaus are common in both maps. A run of three or more consecutive added electrodes that removes nothing occurred in 35 of the 54 sweeps in at least one of the two maps, with the longest run 11; counted per graph this is 23 of the 27. On a separate 47-node network of mean coordination 4.94 (Fig. 3c), five consecutive electrodes between 15 and 20 accessible nodes leave the rank at 112 of 116. An electrode added to an already resolved component removes nothing. The decomposition theorem says so for the scalar map; the same statement for the elastic map is not proved here, and the identification of the plateaus with runs of such electrodes was not tested.

The allosteric response is the displacement at a target under a strain at a source. It is a function of the boundary stiffness map. Two networks on the same fiber of that map therefore have identical allosteric responses at every source and target pair drawn from the accessible set, and a step along a kernel direction changes the response only at second order. Allostery can be trained and measured without the stiffnesses being determined.

\begin{figure*}[t]
\centering
\includegraphics[width=\textwidth]{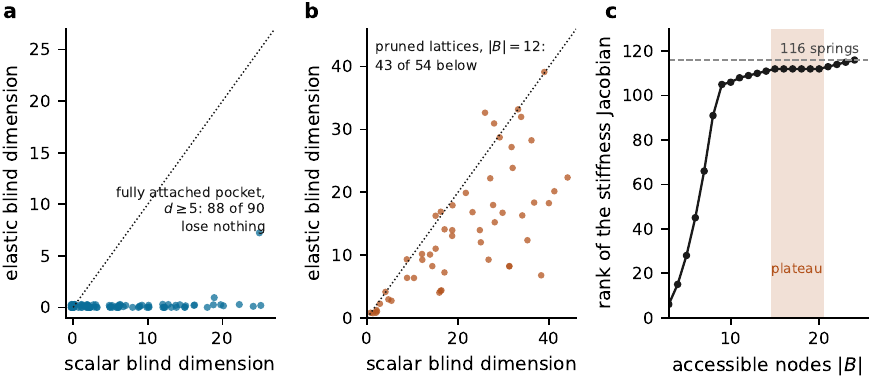}
\caption{\label{fig3}\textbf{a}, Blind dimension of a hidden pocket joined to every accessible node, read as scalar potentials against the same graph read as vector displacements, for the 90 test networks with $d\ge5$; the elastic map loses nothing in 88 of them while the scalar map of the identical wiring averages 6.87. \textbf{b}, The same comparison over the pruned-lattice ensemble at twelve accessible nodes: the elastic blind dimension is smaller in 43 of the 54 sweeps, medians 12.5 against 20.5. \textbf{c}, Rank of the boundary stiffness Jacobian against the number of accessible nodes for one 47-node, 116-spring network of mean coordination 4.94; five consecutive electrodes between 15 and 20 nodes leave the rank at 112.}
\end{figure*}

\section{A tropical lower bound and a conjecture}

For a general hidden component the rank of \(\Pi J_C\) is not given by counting. By the all-minors matrix-tree theorem~\cite{chaiken1982}, every off-diagonal entry of \(\Lambda_B\) is a ratio of two spanning-forest polynomials with unit coefficients. The denominator runs over forests in which each tree contains exactly one accessible node, the numerator over forests in which every tree contains exactly one accessible node, except one tree that contains exactly a and b. Kenyon and Wilson used the same representation to read boundary connection probabilities off the response matrix~\cite{kenyon2011}. Assign generic weights \(w\). Let \(T^*(w)\) and \(F^*_{ab}(w)\) be the maximum-weight forests of the two families (Fig. 4a), and let \(Q(w)\) be the matrix with rows \(\chi_{F^*_{ab}}-\chi_{T^*}\). Its entries are \(-1\), 0 or 1, and they depend only on the ordering of forest weights. The rank meant here is the ordinary rank, not the tropical rank of Develin, Santos and Sturmfels~\cite{dss2005}, which is a different quantity. Scaling \(k_e=\exp(w_e/\varepsilon)\) with \(\varepsilon\to0\), the logarithmic Jacobian of \(\Lambda_B\) converges to \(Q(w)\), and lower semicontinuity of the rank gives

\begin{equation}\operatorname{rank}\,Q(w)\le\operatorname{rank}\,J_\Lambda\quad\text{for every generic } w\end{equation}

This is proved in Appendix~\ref{app:tropical}. The form used in the tests below is the restricted one, on the surviving slots and with the \(E_{BB}\) columns dropped, and the prediction scored on the held-out set is \(\dim\ker J_\Lambda\le(M-|E_{BB}|)-\max_w\operatorname{rank}\,Q_{\mathrm{res}}(w)\), conjectured to be an equality in general. For a component with one hidden node that equality is proved. There the maximum-weight forest of each family is unique, so \(Q(w)\) is the unsigned incidence matrix of F minus a rank-one term. Its rank is \(d-b(F)\) for every ordering, and that is the rank of \(\Pi J_C\), the local Jacobian on the surviving slots. Inequality (5) instantiates Draisma's tropical lower bound on the dimension of the image of a parametrization~\cite{draisma2008}, with grounded spanning forests as the monomials of maximum weight. Inequality and conjectured equality are the two statements of tropical implicitization~\cite{sty2007,st2008}, whose equality theorem holds when the coordinate polynomials are generic for their Newton polytopes. The forest polynomials have all coefficients equal to one, which is the opposite of generic, so the equality does not follow from that theory and a proof would have to use the forest structure. The conjecture was tested in two stages. On the 5,343 hidden components of one to seven hidden nodes carrying two to twelve conductances, forests were enumerated exhaustively and the maximum over orderings was certified rather than searched. Inequality (5) held in every case and equality as well, 5,343 of 5,343 (Fig. 4b). A single ordering attaining the proved cap \(\operatorname{rank}\,\Pi J_C\) settles a component outright, and such a witness was found for every one, 5,339 within sixty draws and the rest within six hundred, each re-verified in exact integer arithmetic with its dominant forests checked to be unique, a tie being a degenerate cell the theory excludes. The 4,606 components carrying nine conductances or fewer, and 26 more at ten, were in addition enumerated over every one of their \(m!\) orderings, the maximum equaling \(\operatorname{rank}\,\Pi J_C\) in 4,632 of 4,632. The equality is not vacuous: 2,496 components have a nonzero blind dimension, and 918 have maximum tropical rank strictly below the counting cap. The best of sixty random orderings instead reaches equality in 5,339, and which four it misses depends on the seed, so those four are failures of the search. For the 2,812 single-node components one ordering suffices in every one, as the proof gives. Among the 2,531 larger components one ordering suffices in 1,579. On a held-out set of 1,500 networks generated with a separate seed and labeled only after the prediction was fixed, the maximum over orderings of the rank of \(Q(w)\) was searched by hill-climbing at the pre-specified budget of 300 evaluations or 60 seconds per hidden block. The blind dimension predicted from forests alone equaled the certified one in 478 of 500 lattices, 488 of 500 random and planar graphs, and 482 of 500 structured ones, 96.5\% overall, clearing the pre-specified threshold of 95\% in each family. The run first reported ran at a 90 second cap that was not pre-specified and gained three lattices, 96.7\% (Appendix~\ref{app:prespec}). In the 52 remaining cases the forest rank fell below the certified one and in none did it exceed it (Fig. 4c); at the 90 second cap there are 49. A shortfall is what an incomplete search produces, and it is also what a false equality would produce, so these do not separate the two by themselves. Re-running the 49 at a raised budget closes 23, raising the agreement to 98.3\% and leaving 26 open (Appendix~\ref{app:prespec}). Each of those 26 traces to a single block, the component deficiencies summing to the certified blind dimension in 26 of 26. Those blocks carry 29 to 68 conductances, putting enumeration thirty orders of magnitude and more out of reach, so they stay undecided. Where a witness can be found at all, the equality has never failed; what separates these blocks from the small components is the size of the search space. Subject to the conjecture, which is open for components with two or more hidden nodes, the blind dimension can be computed from the wiring and the electrode set by a forest count alone, with no linear algebra.

\begin{figure*}[t]
\centering
\includegraphics[width=\textwidth]{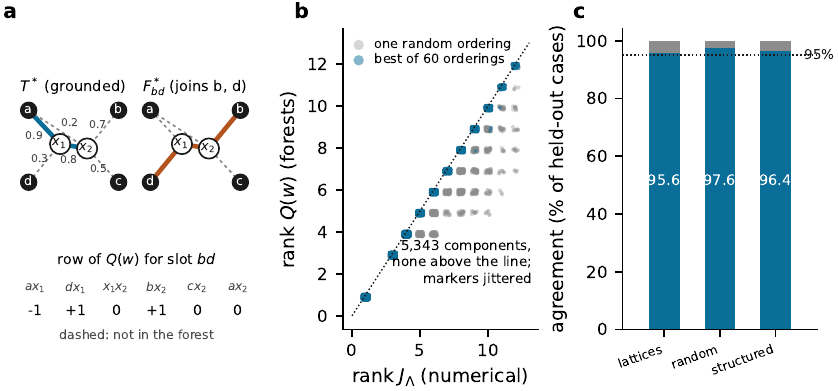}
\caption{\label{fig4}\textbf{a}, On a small hidden component with accessible nodes a to d and hidden nodes $x_1$ and $x_2$, the maximum-weight grounded forest $T^*$ and the maximum-weight forest joining b and d, for one weight ordering; below them the corresponding row of $Q(w)$. Edge weights are marked on the left panel; dashed edges are the ones outside the forest. \textbf{b}, Rank of $Q(w)$ against the rank of $\Pi J_C$, the local Jacobian on the surviving slots, for the 5,343 hidden components of one to seven hidden nodes carrying between two and twelve conductances: no point lies above the diagonal, and the maximum over orderings reaches it in all 5,343, each attained by a certified witness ordering. The best of sixty random orderings instead reaches it in 5,339, a single ordering in 4,391. Markers are jittered along the axis and downward only, so the jitter cannot carry a point across the diagonal. \textbf{c}, Held-out evaluation on 1,500 networks labeled after prediction, at the pre-specified search budget: agreement in 95.6\% for lattices, 97.6\% for random and planar graphs, and 96.4\% for graphs with structured hidden components; the gray remainder is the cases where the forest rank fell below the numerical one. Re-running the 49 disagreements of the 90 second run at a raised budget closes 23 of them, raising the agreement there to 98.3\%; the remaining 26 each trace to a single block of 29 to 68 conductances, too large to enumerate, and are consistent with an incomplete search without ruling out a false equality.}
\end{figure*}

\section{Where to place the next electrode}

When the surviving slot sets of the components are pairwise disjoint, the theorems make electrode placement a local combinatorial question. The blind dimension falls as accessible nodes are added (Fig. 5a), and how fast depends on which node is added (Fig. 5b). On the lattice of Fig. 1a with the deficient eight-node set, each hidden node was made accessible in turn. Five of them remove all seven blind directions, two remove four, and one removes three (Fig. 5b). Reproducing these eight gains from the forest characterization alone was a pre-specified criterion. The characterization reproduces them at some search seeds and not at others: at the deposited default it errs by one on three of the eight nodes, always by reporting a blind direction where there is none. The gains quoted here are computed from \(J_\Lambda\). Which node is added matters: when the surviving slot sets of all components are pairwise disjoint, two electrodes on different components remove the sum of their gains, while a second electrode on the same component can remove nothing.

Both where the electrodes sit and the topology of the network set the deficiency. Over 1,839 networks from thirteen graph families (Appendix~\ref{app:survey}), the excess of the counting bound over the rank was compared across placements of the accessible set. The excess splits into a structural part, which is the counting bound minus the number of slots that can be nonzero at all, capped by the parameter count, and a residual (Appendix~\ref{app:survey}). The excess grows steeply with the number of accessible nodes, so every comparison quoted here is between populations matched on that number and on the graph (Appendix~\ref{app:survey}). Table I gives the three planar families pooled, at three placements compared over the same 95 pairs of graph and accessible-set size. Spreading the electrodes around the outer face in circular order leaves almost nothing: an excess in 4.2\% of cases, and no structural part in any of them. Clustering the same electrodes on one arc is the worst of the three, at a mean excess of 8.69. Placing them in the interior is intermediate, at 3.22, and interior sets keep every slot occupied in 92 of 95 cases. What separates circular order from the interior is the residual; against an arc both parts separate them. Circular order is the condition that Curtis, Ingerman, and Morrow require~\cite{cim1998}. The last two rows are a second and separate comparison, on square lattices read at random accessible sets. The periodic lattice is the easier one to read, at a mean excess of 0.14 against 5.37. It is not the harder one in any of the 59 matched cells, and what excess it keeps is structural with no residual. A change of topology at fixed placement therefore moves the excess by 5.23, where the best and the worst of the three placements differ by 8.51. Those two figures come from different blocks of the table and are not directly comparable. The two blocks rest on different populations, so together they do not rank placement against topology. On the square lattices alone, over the 48 pairs of lattice size and accessible-set size where all five conditions were run, they can be ranked. In the plane the mean excess is 0.25 in circular order, 5.76 at random placement, 6.21 in the interior and 16.98 on an arc, against 0.11 at random placement on the torus. Placement spans about three times the range that the change of topology does there, and neither is negligible.

Circular-planar networks that Curtis, Ingerman, and Morrow call critical, meaning that no local transformation reduces them further, are exactly the ones their theorem recovers in full. On seven such networks, built here by their reduction procedure (Appendix~\ref{app:survey}), the rank equaled M in every case, which is the check that the survey and their theorem agree.

\begin{table*}[t]
\caption{Excess of the counting bound over the rank of $J_\Lambda$, by where the accessible nodes are placed. The table holds two matched comparisons. Rows are comparable within a block and not across the two. The first three rows are three placements of the accessible set, each pooling the three planar families of the survey over the same 95 pairs of graph and accessible-set size, namely 48 square lattices in the plane, 27 random planar and 20 circular-planar networks per row. The last two rows are square lattices read at random accessible sets, in the plane and with periodic boundaries, over the same 59 pairs of lattice size and accessible-set size, two networks per pair. Matching on the accessible-set size is needed because the excess grows steeply with it. The excess splits into a structural part and a residual (Appendix~\ref{app:survey}). Means are over the $n$ networks of each row.}
\label{tab1}
\begin{ruledtabular}
\begin{tabular}{lrrrrr}
placement & n & mean excess & structural & residual & excess $>0$ (\%) \\
\colrule
circular order on the outer face & 95 & 0.18 & 0.00 & 0.18 & 4.2 \\
clustered on one arc of the outer face & 95 & 8.69 & 3.29 & 5.40 & 38.9 \\
interior & 95 & 3.22 & 0.19 & 3.03 & 15.8 \\
random, square lattices in the plane & 118 & 5.37 & 0.67 & 4.70 & 39.8 \\
random, the same lattices with periodic boundaries & 118 & 0.14 & 0.14 & 0.00 & 1.7 \\
\end{tabular}
\end{ruledtabular}
\end{table*}

The noise-free rank is an upper bound on what can be read in practice. A conductance perturbation of relative size one is detectable at relative boundary precision \(\varepsilon\) when the corresponding singular value of the relative Jacobian \(J_\Lambda\,\mathrm{diag}(k)/\|\Lambda_B\|_F\) exceeds \(\varepsilon\), so it is that matrix, and not \(J_\Lambda\) itself, whose spectrum sets the count. On the lattice of Fig. 1 its median spectrum over 20 conductance draws spans 4.1 decades; at a relative measurement precision of \(10^{-2}\) 13 of the 25 identifiable directions are above the floor, at \(2\times10^{-3}\) 19, and all 25 once the precision is finer than the smallest nonzero singular value, \(4\times10^{-5}\) (Fig. 5c). The counts depend on the scaling: the median spectrum of \(J_\Lambda\) itself over the same 20 draws spans 3.5 decades and gives 22, 24 and 25. The method follows Ref.~\cite{boyer2016}; the spread of the spectrum over decades is the sloppiness of Gutenkunst and co-workers~\cite{gutenkunst2007} in this setting. The numbers are specific to this network and are meant as a scale, not as a hardware specification.

\begin{figure*}[t]
\centering
\includegraphics[width=\textwidth]{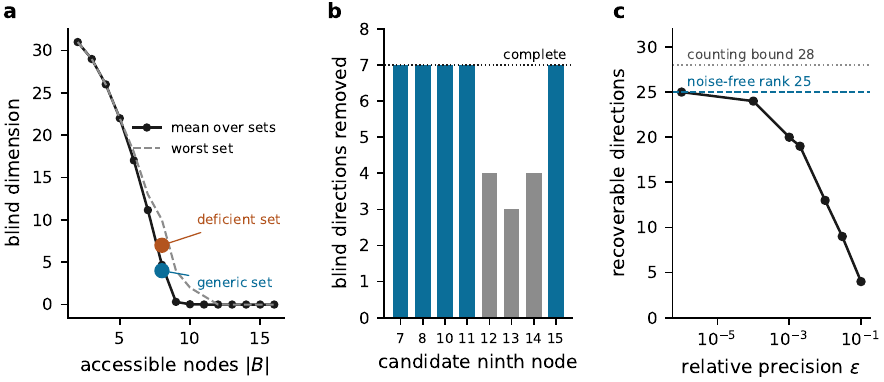}
\caption{\label{fig5}\textbf{a}, Blind dimension against the number of accessible nodes on the $4\times4$ periodic lattice: mean over random sets of each size, and the worst set of each size, with the generic and the deficient eight-node set marked. \textbf{b}, Reduction of the blind dimension when each of the eight hidden nodes is exposed, starting from the deficient set. The node indices on the axis are those of the lattice of Fig. 1a. Five nodes remove all seven directions and are drawn in color; two remove four and one removes three. The forest characterization reproduces these gains at some search seeds and errs by one on three nodes at the deposited default. \textbf{c}, Number of recoverable directions against relative measurement precision for the same network, from the median singular spectrum of the relative Jacobian $J_\Lambda\,\mathrm{diag}(k)/\|\Lambda_B\|_F$ over 20 conductance draws; the noise-free rank is 25 and the counting bound 28.}
\end{figure*}

\section{Visible and blind directions in a self-learning circuit}

The split between visible and blind directions can be read from published hardware trajectories: the gate voltages of all 32 self-adjusting resistors and the voltages of all 16 nodes at 42 measurement steps, for 190 regression tasks on a contrastive local learning network~\cite{dillavou2024,dryad2024}. The metadata give a \(4\times4\) lattice with periodic connections in both directions, so the graph is the torus of Fig. 1; the edge-indexing convention was fixed by Kirchhoff residuals (Appendix~\ref{app:hardware}). Each task drives three source nodes and reads one target, so \(|B|=4\), six slots, rank 6, and 26 blind directions of 32. With four accessible nodes the rank saturates the counting bound at every step, so this data set shows the split between visible and blind directions and does not test the combinatorial results above.

Conductances were reconstructed from the gate voltages with the transistor law of Ref.~\cite{dillavou2024}, and each step's conductance increment was projected onto \(\ker J_\Lambda(B)\) evaluated at the mean of the two consecutive states it joins. For an ideal linear network the projection would vanish, because the contrastive function that coupled learning descends is a function of \(\Lambda_B\) and its gradient is orthogonal to the blind subspace; the measured fraction is the departure from that ideal. Such a fraction depends on the coordinates the increment is written in, so it was computed twice, once for the conductance increment against \(\ker J_\Lambda\) and once for the relative increment against the kernel of \(J_\Lambda\) diag(k), which is the same statement in log-conductance coordinates. Table II gives the four numbers, over the 178 experiments that have both a learning phase and a plateau. While the error falls, the fraction is below the isotropic value in all 178 experiments in both coordinate systems (Fig. 6a, b for conductance coordinates; Table II for both), so the bias toward visible directions during learning is not an artifact of the coordinates.

\begin{table*}[t]
\caption{Fraction of each conductance update that lies in the blind subspace, for the 178 hardware experiments that have both a learning phase and a plateau. Each row is the median over experiments of the per-experiment median; the isotropic column is the expectation for an update with no preferred direction, computed on the edges that moved. The learning-phase result holds in both coordinate systems and the plateau result does not.}
\label{tab2}
\begin{ruledtabular}
\begin{tabular}{llrrr}
phase & coordinates & median fraction & isotropic & below isotropy \\
\colrule
learning & conductance & 0.344 & 0.812 & 178 of 178 \\
learning & log conductance & 0.375 & 0.812 & 178 of 178 \\
plateau & conductance & 0.707 & 0.849 & 168 of 178 \\
plateau & log conductance & 0.865 & 0.786 & 53 of 178 \\
\end{tabular}
\end{ruledtabular}
\end{table*}

The plateau is where the two coordinate systems part. In conductance coordinates the fraction rises to 0.707 against an isotropic 0.849 and stays below it in 168 of 178 experiments, which reads as a visible bias surviving at 0.30 of its size during learning. In log-conductance coordinates the same experiments give 0.865 against 0.786, below in 53 of 178, and the sign of the deviation reverses. The residual plateau bias is therefore a property of the coordinates, and it is not reported here as a result. The learning-phase bias survives both coordinate systems, and that one is reported. No significance is attached to any of these counts: the 190 tasks ran on one device with one calibration, so they are not independent replicates. The boundary response itself keeps changing in the plateau, by a relative amount per step close to that of the conductances (median over plateau steps of the ratio 1.04, Fig. 6c), although the task error is flat: the trained network moves in directions the task does not check. A pre-specified test of the stronger statement, that the plateau drift concentrates in the blind subspace beyond isotropy, was not confirmed in any of the four combinations of subspace and plateau definition that the record fixes. All four are in conductance coordinates, and there the deviation runs the other way. In log-conductance coordinates the plateau fraction lies above the isotropic value in 125 of 178 experiments, which is the direction the prediction asserted, so the outcome of the pre-specified test is itself coordinate-dependent (Appendix~\ref{app:hardware}).

\begin{figure*}[t]
\centering
\includegraphics[width=\textwidth]{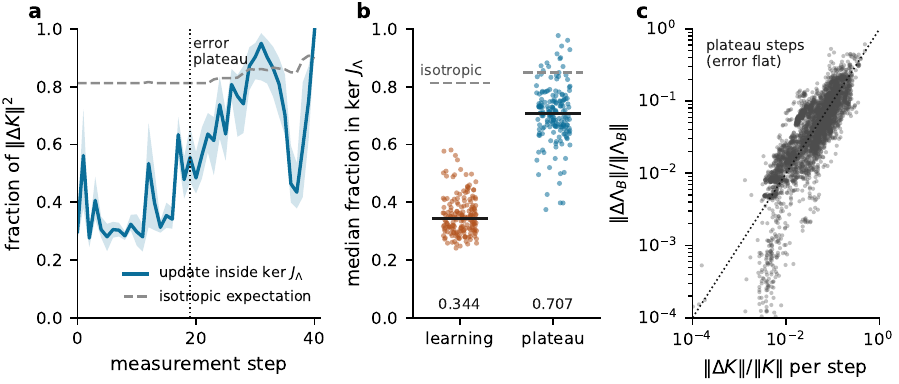}
\caption{\label{fig6}Data of Ref.~\cite{dillavou2024} (190 regression tasks on a $4\times4$ periodic lattice, $|B|=4$). \textbf{a}, Median fraction of the conductance increment that lies in $\ker J_\Lambda(B)$ at each measurement step (line, interquartile band) against the isotropic expectation (dashed); the dotted line marks the median onset of the error plateau. \textbf{b}, Per-experiment medians in the learning and plateau phases over the 178 experiments that have both: 0.344 and 0.707, against isotropic expectations of 0.812 and 0.849. Both panels are in conductance coordinates; Table II gives the same four quantities in log-conductance coordinates as well. \textbf{c}, Relative change of $\Lambda_B$ per plateau step against the relative change of the conductances (median ratio 1.04): the boundary response keeps moving while the task error is flat.}
\end{figure*}

\section{Discussion}

The blind dimension of a learning network is a property of its wiring and of its electrode set at generic conductances. Four theorems about the blind dimension are proved here, for linear resistor networks read at a fixed operating point. One of them, the exposure bound, is carried over in the same shape to central-force spring networks, under the added hypothesis that the pocket is rigid when its boundary neighbors are pinned; the other three have no elastic analogue here. The blind dimension decomposes over hidden components whose surviving slots are disjoint, and it is never decreased by a boundary-to-boundary edge. A component with a single hidden node contributes a closed-form amount. What electrodes outside a component of any size can expose is bounded by a factor-analysis count. That count saturates as electrodes are added around a completely attached pocket with no edge between the pocket's neighbors. An electrode inside such a pocket, when its hidden nodes form a clique, exposes \(h-1\) directions. A bound of the same shape holds for the elastic reading. And a forest count bounds the blind dimension from above, which is inequality (5); that bound is an equality for a component with one hidden node. Two further statements are not theorems. The forest bound is conjectured to be an equality for components with two or more hidden nodes as well. On the 5,343 components whose maximum over weight orderings could be attained and certified, the equality holds without exception, which is the strongest evidence short of a proof; on held-out networks, where the search space is far larger, it was matched in 96.5\% of cases at the pre-specified budget. And on the same wiring the elastic map left no deficit beyond counting in all 90 spring networks tested, each a pocket joined to every one of five or more accessible nodes, which is an observation on those 90. For the transistor network of Ref.~\cite{dillavou2024} all of this applies to the linearization about the current state, which is how it was used in Fig. 6.

The claims are bounded as follows. The tropical equality is a conjecture, open for components with two or more hidden nodes; the maximum is certified on every component whose search space is small, leaving open exactly those whose search space is not. The forest prediction also falls short of the certified value in 3.5\% of held-out cases at the pre-specified budget, 3.3\% at the 90 second cap actually used first, and 1.7\% at the larger one. The protocol set two acceptance criteria and the second was not met: the forest characterization does not reproduce the eight electrode gains of Fig. 5b at the deposited search seed. A first held-out run was discarded for a genericity fault in its weights, and it had scored better than the run reported here on both criteria. Both runs used a 90 second per-block time cap where 60 seconds was pre-specified, a deviation that runs in the direction of higher agreement (Appendix~\ref{app:prespec}). A shortfall does not distinguish an incomplete search from a false equality. The exposure bound is proved sharp only under complete attachment. Under partial attachment it was attained in 415 of 564 test cases. It is also the binding term in only 1 of 400 held-out networks, where it moves the bound from 33 to 32 against a true rank of 30. The elastic bound's determinantal term is loose and never binds at two dimensions, so the elastic tests exercise the parameter count and the rigid-motion cap rather than the new term. The advantage of vector over scalar reading also decays as accessible nodes are added. At twenty nodes the elastic blind dimension is the smaller one in only 14 of 41 sweeps, though the medians there are still 8 against 9. Ranks in the held-out ensemble and in the elastic networks are certified: all 1,500 and all 120 were recomputed in exact rational arithmetic and over \(\mathrm{GF}(p)\) and agree with the singular value decomposition throughout. Those in the exploration ensemble and in the 12,870 accessible sets still come from singular value decomposition, checked exactly on 200 cases. On the placement survey the singular value decomposition undercounted the rank in 12 of 36 flagged rows, by 19 in one of them. The theory counts noise-free directions: on the lattice of Fig. 1 only 13 of the 25 identifiable directions are above the floor at a relative measurement precision of \(10^{-2}\). The hardware comparison is descriptive. Its conductances are reconstructed rather than measured, with a Kirchhoff residual near 15\%, and the stronger drift statement that was pre-specified failed in the conductance coordinates it was written in, while the same quantity in log-conductance coordinates runs the other way. The residual bias in the plateau is present in conductance coordinates and absent in log-conductance coordinates, so only the learning-phase bias is claimed. Non-central forces, prestress, and bending, present in the elastic experiments of Refs.~\cite{pashine2019,rocks2017,rocks2019}, change the boundary map and were not treated.

For the design of learning matter the practical content is a rule. Before a training campaign, the hidden components of the wiring and the slots deleted by boundary edges fix how many directions training at B can ever set, and when the components occupy disjoint slots the gain of each candidate electrode can be evaluated on its own component. The count is a dimension and not a list: which conductances are pinned does not follow from it, as the three-conductance example of the decomposition section shows. The saturation law, which is Ledermann's count read through the electrical parametrization, says where the evaluation ends: a completely attached pocket of \(h\) hidden nodes with no edge between its neighbors, already joined to \(d^*(h)\) accessible nodes, will not expose more to any electrode placed outside it. Breaking such a pocket up, when its hidden nodes form a clique, by making one of them accessible, is worth \(h-1\) directions where surrounding it is worth none. The remaining option is to measure more per node: on the same wiring the elastic map of a completely attached pocket with five or more accessible neighbors left no deficit beyond counting in all 90 networks tested. Below five neighbors the elastic map was deficient too, and the counting value held in only 15 of 30. For the interpretation of trained states, the fiber of Fig. 1b sets a limit: a walk along it moved one conductance to 15.2 times its starting value while every response at the accessible nodes stayed fixed, so a network read after training is not pinned by what the supervisor can see. Local rules that read only the accessible nodes are now built as boundary-driven static schemes~\cite{ezraty2026} and as time-domain adjoint drives on mechanical networks~\cite{limao2026}, and both act through a response map; how many boundary measurements a graph forces is a separate budget, worked out for locating a faulty edge from effective resistances~\cite{fiedorowicz2026}. Whether the same decomposition organizes the flat directions of nonlinear and of dynamic learning networks, and whether the forest characterization can be turned into a closed formula for hidden trees and lattice clusters, are the next questions.

\textbf{Acknowledgments.} The authors thank the authors of Ref.~\cite{dillavou2024} for depositing their raw trajectories.

\textbf{Author contributions.} Q.-B.N. wrote the code, ran the ensembles, and analyzed the data. T.-S.V. conceived the study, proved the results, and wrote the paper. Both authors checked the numerical results and approved the manuscript.

\textbf{Competing interests.} The authors declare no competing interests.

\textbf{Use of generative AI.} A generative AI assistant (Anthropic Claude) was used to assist with language editing and selected computational tasks. The authors formulated the research questions and mathematical arguments, reviewed and verified the references and numerical results, and take full responsibility for the content of this work.

\appendix

\section{Response map and Jacobian}\label{app:jacobian}

For a connected graph with weighted Laplacian L, \(\Lambda_B=L_{BB}-L_{BI}L_{II}^{-1}L_{IB}\). With \(P\) the harmonic-extension matrix, whose rows on B form the identity and whose rows on I are \(-L_{II}^{-1}L_{IB}\), the drop vector of edge \(e=(i,j)\) is \(u_e=P_i-P_j\). Writing \(\Lambda_B=P^{\mathsf T}LP\) and differentiating, the two terms in \(\partial P\) drop out, because \(LP\) has zero rows on the hidden set while \(\partial P/\partial k_e\) has zero rows on the accessible set, which leaves Eq. (1). \(J_\Lambda\) was assembled as the matrix with rows indexed by slots \(\{a,b\}\), a \textless{} b, and entries \(u_e[a]u_e[b]\); its rank at generic conductances, drawn log-uniformly over two decades, was computed by singular value decomposition with tolerance \(\sigma_{\max}\cdot\max(\text{shape})\cdot\varepsilon_{\text{machine}}\) and checked on three independent draws. Two hundred cases drawn at random from the two ensembles were recomputed in exact rational arithmetic (python-flint) at random integer conductances, two independent draws each: the two exact ranks agreed in all 200, and in the 59 that carry a held-out label the exact blind dimension equaled that label. All 1,500 evaluated held-out cases were then certified rather than sampled: recomputed in exact rational arithmetic at three random integer conductance draws and over \(\mathrm{GF}(p)\) for six primes at four draws, with eight further modular draws as a check. Modular and rational agreed in all 1,500, so these carry the stronger rational certification, and no draw ever raised the rank. The certified rank equals the singular value decomposition rank in 1,500 of 1,500, so every held-out figure here rests on certified ranks. All 1,500 carry a clean singular-value gap, so the ensemble sits in the class the flag never catches, and none was wrong, the 95\% upper bound on that silent error rate being 0.20\%. That the certification catches undercounts was checked on a control of 144 lattice networks reaching \(M=288\), where the singular value decomposition undercounts in eight, none below \(M=100\); the held-out ensemble reaches \(M=104\). The 120 elastic configurations were certified the same way and agree in 120 of 120. An exact rank at a sampled conductance is a lower bound on the generic rank and equals it off a proper closed subset, so where the word exact is used below it means the largest rank found in exact arithmetic over the draws performed, not a proof that no larger rank occurs. The script is deposited.

\section{Level sets}\label{app:levelsets}

The fiber through \(k_0\) was followed by a predictor step of length 0.02 along a kernel direction and a Gauss--Newton corrector onto \(\{\Lambda_B=\Lambda_B(k_0)\}\), with the tangent direction propagated by projection onto the current kernel to keep it continuous. The walk stopped when a conductance reached zero.

\section{Proof of the decomposition theorem}\label{app:decomp}

Because there are no edges between different hidden components, \(L_{II}\) is block diagonal and \(L_{BI}L_{II}^{-1}L_{IB}=\sum_C L_{BC}L_{CC}^{-1}L_{CB}\), each summand supported on \(N(C)\times N(C)\); this is Eq. (2). The column of \(J_\Lambda\) for a boundary edge ab is \(-e_{ab}\) in slot coordinates, since \(k_{ab}\) enters no other off-diagonal entry. For a matrix whose first block consists of distinct coordinate vectors, the rank is the number of those vectors plus the rank of the second block projected off the coordinates they occupy. If the surviving slot sets of the components are pairwise disjoint, \(\Pi[J_{C_1}\cdots J_{C_p}]\) is block diagonal up to a permutation and its rank is the sum of the block ranks. Exposing \(v\in C\) replaces C by the components of \(C\setminus\{v\}\) and moves the edges at v into \(E_{BB}\); no other block is touched.

\section{Proof of the one-node formula}\label{app:onenode}

With \(k_a\) the conductance of the edge from \(x\) to \(a\), \(\kappa=\sum_c k_c\), and \(p=k/\kappa\), the surviving slots are the edges of F and \(S^C_{ab}=k_ak_b/\kappa\). In logarithmic coordinates the Jacobian has rows \(e_a+e_b-p\), that is, \(R=V-\mathbf 1p^{\mathsf T}\) with V the unsigned incidence matrix of F. Since \(V\mathbf 1/2=\mathbf 1\), \(R=V(I-zp^{\mathsf T})\) with \(z=\mathbf 1/2\), and \(\det(I-zp^{\mathsf T})=1-p^{\mathsf T}z=1/2\ne0\), so \(\operatorname{rank}\,R=\operatorname{rank}\,V=d-b(F)\)~\cite{grossman1995}. Positive diagonal rescalings do not change rank, so the same holds for the Jacobian in conductance coordinates.

\section{Proof of the exposure bound}\label{app:exposure}

\(L_{CC}\) is a principal submatrix of an irreducible weighted Laplacian with at least one row and column deleted, hence positive definite, so \(S^C=L_{N(C)C}L_{CC}^{-1}L_{CN(C)}\) is positive semidefinite and factors through \(\mathbb R^h\), giving \(\operatorname{rank}\,S^C\le h\). Let \(\Sigma_{d,h}\) be the positive semidefinite matrices of rank at most \(h\) in \(\mathrm{Sym}_d\) and \(\pi\) the map that drops the diagonal. For \(h\ge d\), adding a multiple of the identity puts any symmetric matrix in \(\Sigma_{d,h}\) without changing its off-diagonal part, so \(\pi(\Sigma_{d,h})\) is everything. For \(h<d\), the factor-analysis variety \(\mathbf F_{d,h}=\Sigma_{d,h}+\mathrm{Diag}_d\) has \(\dim\mathbf F_{d,h}=\min\{d(h+1)-h(h-1)/2,\ d(d+1)/2\}\)~\cite{dss2007}; \(\pi(\mathbf F_{d,h})=\pi(\Sigma_{d,h})\), and every fiber of \(\pi\) on \(\mathbf F_{d,h}\) is a full translate of \(\mathrm{Diag}_d\), so \(\dim\pi(\Sigma_{d,h})=\dim\mathbf F_{d,h}-d=\varphi(h,d)\). At generic \(k\) the rank of the Jacobian of a rational map is the dimension of its image, and the image here is a coordinate projection of \(\pi(\Sigma_{d,h})\); splitting the surviving slots into those contained in \(N_0\) and the rest, and using that the principal submatrix of \(S^C\) on \(N_0\) is again positive semidefinite of rank at most \(h\), gives the bound. For sharpness under complete attachment, set the internal conductances of the pocket to \(\varepsilon c\) and let \(\varepsilon\to0\). Then \(L_{CC}\) becomes diagonal, \(S^C=\sum_v a_va_v^{\mathsf T}/|a_v|_1\) with \(a_v\) the vector of attachment conductances of \(v\), and the substitution \(b_v=a_v/\sqrt{|a_v|_1}\) is a bijection of the positive orthant, so the image is that of \(B\mapsto B^{\mathsf T}B\) on positive \(B\). Here \(B\) is the \(h\times d\) matrix whose rows are the \(b_v\). The differential of \(B\mapsto B^{\mathsf T}B\) at a full-rank \(B\) has kernel \(\{AB:A^{\mathsf T}=-A\}\). For \(h\le d\) that kernel has dimension \(h(h-1)/2\), so the rank is \(hd-h(h-1)/2\). For \(h\ge d\) the rows of \(B\) span \(\mathbb R^d\), which forces the image of \(A\) into an \((h-d)\)-dimensional subspace, so the kernel drops to \(h(h-1)/2-(h-d)(h-d-1)/2\) and the rank is \(d(d+1)/2\). In both cases the image contains a relatively open piece of the smooth locus of \(\Sigma_{d,h}\), and \(\pi\) of it has dimension \(\varphi(h,d)\). Rank is lower semicontinuous, so the value persists for small \(\varepsilon>0\), and the rank is generically constant in \(k\). With complete attachment \(m_C=|E_{\mathrm{int}}(C)|+hd\) and \(\varphi(h,d)=hd-h(h-1)/2\) whenever \(d^2-(2h+1)d+h^2-h\ge0\), that is \(d\ge d^*(h)\), which gives equation (4). Exposing \(t\) nodes of a clique pocket deletes every slot that meets an exposed node, so \(N_0=N(C)\) is the best subset and the same argument gives \(\delta=(h-t)(h-1)\).

\section{Why the elastic determinantal term never binds}

The pocket is assumed rigid when its neighbors are pinned, so the \(D_r\) rigid motions of \(C\cup N(C)\) restrict injectively to \(N(C)\) and \(Dd\ge D_r\). Under that hypothesis \(\psi(Dh,Dd)\ge\min\{m_C,\ (Dd-D_r)(Dd-D_r+1)/2\}\) for every \(D\ge2\), so the middle term of the elastic bound is never the unique minimum and the bound reads \(\min\{m_C,\ (Dd-D_r)(Dd-D_r+1)/2\}\). For \(h\ge d\) this is the monotonicity of \(t\mapsto t(t+1)/2\) at \(t=Dd-D_r<Dd\); for \(h<d\) it follows from \(m_C\le hd+h(h-1)/2\), since \(\psi-m_C\) is then bounded below by a bracket decreasing in \(h\) whose value at \(h=d\) is \(\{d(D^2-3)+(D+1)\}/2>0\). The hypothesis is not cosmetic: at \(D=4\), \(d=1\), \(h=5\) a clique pocket gives \(\psi=10\) below both other terms, and there a rotation about the single neighbor is a zero mode of the pinned pocket, so the hypothesis fails exactly where the conclusion does. Across the 120 elastic networks \(\psi\) attains the minimum in none, its smallest margin being 8.

\section{Proof of the tropical bound}\label{app:tropical}

By the all-minors matrix-tree theorem~\cite{chaiken1982}, \(\Lambda_{ab}=-\big(\sum_{F\in\mathbb F_{ab}}k^F\big)/\big(\sum_{T\in\mathbb F}k^T\big)\) with \(\mathbb F\) the grounded spanning forests, in which every tree contains exactly one accessible node, and \(\mathbb F_{ab}\) those in which one tree instead contains exactly a and b. Let \(r\) be the generic rank of \(J_\Lambda\). Because the rank is lower semicontinuous and constant on a dense open set, r is also its maximum over all positive conductances, so along the curve \(k_e(\varepsilon)=\exp(w_e/\varepsilon)\) the rank is at most \(r\) for every \(\varepsilon\), with no condition on \(w\). In logarithmic coordinates the entries of the Jacobian of \(\log|\Lambda_{ab}|\) in \(\log k_e\) are the difference of the fraction of the numerator's weight and of the denominator's weight carried by forests through e; as \(\varepsilon\to0\) the maximum-weight forest of each family dominates the rest by \(\exp(\Delta/\varepsilon)\), so those entries converge to \(\chi_{F^*_{ab}}(e)-\chi_{T^*}(e)\), which are the entries of \(Q(w)\). Positive diagonal rescalings do not change rank, and no row vanishes because \(\Lambda_{ab}\) is strictly negative wherever \(\mathbb F_{ab}\) is nonempty. A slot with \(\mathbb F_{ab}\) empty carries no information; it is deleted from the slot set before \(Q(w)\) is formed, in the statement and in the code alike. Lower semicontinuity of the rank then gives \(\operatorname{rank}\,Q(w)\le r\). Distinct forest weights are the only hypothesis, and they are exactly what the implementation tests. The form used below is the restricted one, on the surviving slots and with the \(E_{BB}\) columns dropped. It follows in two steps. No grounded forest contains a boundary-to-boundary edge at all, because each of its trees holds exactly one accessible node, so \(T^*\) never does. For a surviving slot \(ab\) the same holds for \(F^*_{ab}\): its trees hold one accessible node each except the one that holds exactly \(a\) and \(b\), and \(ab\notin E_{BB}\). The \(E_{BB}\) columns therefore vanish identically. The same limit argument applied to the surviving-row and non-\(E_{BB}\)-column submatrix then gives \(\operatorname{rank}\,Q_{\mathrm{res}}(w)\le\operatorname{rank}\,\Pi[J_{C_1}\cdots J_{C_p}]\), which is the quantity the held-out evaluation scores. For a component with one hidden node the maximizer of each family is unique once the slots occupied by boundary edges are deleted, and the argument of the one-node formula gives \(\operatorname{rank}\,Q(w)=d-b(F)=\operatorname{rank}\,\Pi J_C\) for every ordering, which proves the equality in that case.

\section{Pre-specified criteria and deposited checks}\label{app:prespec}

The criteria for accepting the tropical characterization were written down before the held-out set was labeled and are deposited with the code. The pre-specification record carries no third-party timestamp, so the criteria are described throughout as pre-specified rather than registered. They were two: agreement with the numerical blind dimension on at least 95\% of each held-out family, and reproduction of the eight electrode gains of Fig. 5b. The first was met, at 95.6\%, 97.6\% and 96.4\% at the pre-specified cap. The second was not met at the deposited default search seed, where the forest characterization reports one blind direction too many at three of the eight nodes, so those three gains come out one too small. A first held-out run was discarded because its implementation accepted weight orderings with tied forest weights, which are outside the definition of the tropical Jacobian and produced five cases violating inequality (5); the corrected implementation rejects such orderings. That run had already been scored when the fault was found: it stood at 97.3\% and it did reproduce the eight electrode gains, so it scored better on both criteria than the reported one. Its numbers are in the deposited pre-specification record. It was discarded because its tropical Jacobian was not the object the proposition is about, a reason independent of its score. All 49 held-out mismatches were then re-run with the evaluation budget raised from 300 to 20,000 evaluations. That nominal budget was never spent: the re-run raised the evaluation count but left the wall clock at its default, so every block stopped on the clock, and the six smallest performed 388 to 1,504 evaluations, 1.9\% to 7.5\% of the nominal figure. The 98.3\% is what the clock bought, and re-searching those six at a longer clock still raises the tropical rank, so the search is not converged at any budget used here. The pre-specified budget was 300 evaluations or 60 seconds per hidden block, whichever came first. The implementation used a 90 second cap instead of 60, in the main held-out run and in the re-run alike. That change was neither decided nor recorded in advance, and it was found in review after both runs had been scored. The first held-out run is therefore a run at 90 seconds. It has since been repeated at the pre-specified cap, on the same 1,500 cases in the same order with the same seed, and that repeat is the primary figure here: 1,448 of 1,500, 96.5\%, by family 478, 488 and 482 of 500. Twelve cases move between the two caps, all to a higher forest count as less search time must, three crossing from agreement to disagreement and none the other way. Of the 2,012 hidden blocks, 213 are trivial, 1,472 reached the counting cap before either limit, 287 stopped on the counter and 40 on the clock; the per-block record is deposited. The cap is tested between evaluations, so a block runs past it by the length of whichever evaluation is in progress. In the main run 25 of the 1,500 cases ran longer than 60 seconds and 19 longer than 90, the longest at 99.8 seconds. In the re-run 44 of the 49 cases ran longer than 60 seconds and 22 longer than 90, the longest at 102.3 seconds. At the pre-specified cap 33 cases ran past 60 seconds and none past 90, the longest overshoot being 9.6 seconds. The three wall-clock counts are from separate executions run days apart and are not comparable case by case. More search time can only raise the rank the search finds, which lowers the forest count and raises agreement, so the deviation runs in the direction that flatters every held-out figure reported here. Twenty-nine improved, of which 23 reached the numerical value exactly; 19 were unchanged and one came out worse at a different random start; the total overshoot fell from 76 to 47, and in none did the forest rank exceed the numerical one. Counting the 23 as agreements raises the held-out figure from 1,451 to 1,474 of 1,500, that is from 96.7\% to 98.3\%, by family 486, 493 and 495 of 500. The ladder from 96.5\% to 96.7\% to 98.3\% is one of search budget, not of evidence for the equality. Neither figure separates an incomplete search from a false equality, since a shortfall is what both produce. The pre-specified number is the first one; the second is what a larger search budget buys and is reported as such. The small verifications quoted in the Results are deposited as a script that recomputes them: the constancy of the rank over 200 conductance draws, the fourteen fiber walks, and the 1,192 single-node exposures behind the locality statement. The six edge-indexing conventions are compared in the deposited script \texttt{hardware\_metric.py}, under the single residual definition given below. The figures quoted below are the \texttt{conventions} entry of its output file \texttt{hardware\_metric.json}, which is the only deposited value of this quantity. The script that labels the held-out ensemble is deposited separately from the script that predicts it. The tests of the exposure bound are deposited in the same way and were run after the bound was proved: 711 random components, 400 held-out networks, 192 complete-attachment networks, a separate family of 200 networks built to sweep \(d\) at fixed \(h\), of which the 136 that lie in the saturated regime test the saturation law, and the 50 interior-electrode cases. An earlier version of the bound, which used \(hd-h(h-1)/2\) with no case split, is wrong for \(h\ge d\) and was violated in 21 of 384 test cases; the failure and the correction are in the deposited pre-specification record.

\section{Ensembles}\label{app:ensembles}

The exploration set (seed 20260904) and the held-out set (seed 20260905) each contain 4,500 cases, 1,500 from each of three families: (a) grid lattices, planar or periodic, with each side drawn independently from 3 to 6, and random accessible sets of size 3 to \(n-1\); (b) connected Erd\H{o}s--R\'enyi graphs and thinned Delaunay graphs on 8 to 20 nodes with random accessible sets; (c) hidden components of prescribed type (path, cycle, star, clique, random tree, 1 to 6 nodes, one to three components) attached at random to 4 to 10 boundary nodes with random boundary edges. The held-out evaluation used the first 500 cases of each family in a fixed random order, labeled after the predictions were computed. A first evaluation run was discarded because the implementation accepted tied weight orderings; the fault was found through five cases that violated inequality (5), which the corrected implementation removed. A separate family of 960 networks tests how sharp the exposure bound is away from complete attachment. It sweeps \(h\) from 1 to 5, \(d\) from 3 to 8, and the fraction of pocket-to-neighbor edges present at 0.4, 0.6, 0.8 and 1.0, with eight draws of each combination. Of the 960, 396 come out completely attached and 564 partially attached. It is deposited as \texttt{exposure\_sharpness.json}.

\section{Placement survey}\label{app:survey}

The support bound is \(\min(M,N_{\mathrm{eff}})\) with \(N_{\mathrm{eff}}\) the number of slots that are not identically zero, equal to the edge count of the fully Kron-reduced graph; the structural part of the excess is the counting bound minus the support bound, and the residual is the support bound minus the rank. Ranks were computed by singular value decomposition and then recomputed over \(\mathrm{GF}(p)\) at random integer conductances wherever that was affordable; where both exist the modular value is the one used. A rank over \(\mathrm{GF}(p)\) is a lower bound on the rank over the rationals and equals it for all but finitely many primes, so it is not the same guarantee as the rational arithmetic used on those 200 cases, and the two are named separately throughout. The main sweep computed the exact rank only when \(M\le200\), which left 36 of the 66 rows with a non-clean singular-value gap carrying a singular value decomposition rank alone; those 36 were recomputed exactly by replaying the sweep for the eight graphs involved with the same seed, and the replay was checked row by row against the stored table before any rank was used. The singular value decomposition had undercounted in 12 of the 36, by 1 in eight cases, by 2 in two, and by 7 and 19 in one each; the corrected ranks are in the deposit and are the ones quoted above. Those two large undercounts are a warning about the method rather than about these rows. The held-out ensemble and the elastic networks have since been certified outright, all 1,500 and all 120 of them. The exploration ensemble and the 12,870 accessible sets are still ranked by singular value decomposition, checked against exact rational arithmetic on 200 cases, of which 59 carry a held-out label and so allow a comparison. After the correction no row with a non-clean gap is without an exact rank. Twenty-two rows have a rank that varied between conductance draws; all of them now carry an exact rank, which is the value used. The 1,839 networks of the placement survey belong to thirteen families (square lattices in the plane, on a cylinder and on a torus; hexagonal and triangular lattices; random planar, circular-planar, random regular, Erd\H{o}s--R\'enyi, small-world, complete, complete bipartite, and trees), with accessible sets placed at random, spread around the outer face, clustered on an arc of the outer face, or in the interior. The excess grows steeply with \textbar B\textbar, and the arms of the survey do not cover the same \textbar B\textbar{} values, so the rows of Table I are matched populations rather than whole arms. Each comparison keeps the main sweep and the cells that every arm of that comparison reaches. A cell is a graph and a value of \textbar B\textbar{} for the placement rows, and a lattice size and a value of \textbar B\textbar{} for the lattice rows. Within a kept cell each arm carries the same number of replicates, which the script asserts, so every row mean is balanced over the same cells. A cell missing from an arm is missing because that placement does not exist on that graph at that \textbar B\textbar. The matching leaves 95 networks per placement row, with the same family composition of 48 square lattices in the plane, 27 random planar and 20 circular-planar networks in each, and 118 networks per lattice row. Pooled instead over each arm's own sample, the placement means are 0.22 over 116 networks in circular order, 11.01 over 126 on an arc, and 7.17 over 132 in the interior, and those three samples differ in their \textbar B\textbar{} distributions. The plane arm of the lattice comparison carries 20 further rows from a separate run on the \(12\times12\) lattice, six of them at \textbar B\textbar{} = 30, 36 and 44, which the torus arm reaches at no lattice size. Its mean excess is 6.67 over all 138 networks, 5.64 over the 132 at \textbar B\textbar{} values the torus arm also covers, and 5.37 over the 118 matched networks. The torus arm carries the \(4\times4\) periodic lattice twice, once as the generic torus grid and once as the isomorphic reconstruction of the network of Ref.~\cite{dillavou2024}; the matched arm keeps one of the two, which moves its mean excess from 0.12 over 130 networks to 0.14 over 118. The five-arm comparison on square lattices quoted in the Results applies the same rule to the 48 cells that random, circular-order, arc and interior placement in the plane and random placement on the torus all reach. The seven critical circular-planar networks are not taken from Ref.~\cite{cim1998}; they were generated here. Seven random circular-planar graphs with 5 to 8 boundary nodes on the outer face and 4 to 9 interior nodes were built and then reduced to critical form by the delete-and-contract procedure of Ref.~\cite{cim1998}, which leaves the boundary connection pattern unchanged. Their specifications and edge lists are in the deposited script.

\section{Spring networks}\label{app:springs}

Two-dimensional central-force networks were built from a triangular lattice of \(5\times5\) to \(7\times6\) sites with positions jittered by 6\% of the lattice constant, dangling nodes of degree two or less pruned, and bonds then removed one at a time in a fixed random order of the initial edge list, each removal kept only when rigidity survived, down to target mean coordinations of 4.4 and 4.0 (the isostatic value in two dimensions is 4, below which the network loses rigidity~\cite{lubensky2015}). The third coordination of each geometry is the unpruned network, of mean coordination up to 4.85. Rigidity was checked by requiring the stiffness matrix to have exactly three zero modes. Three replicas of each of the three sizes were generated and each was taken at the three coordinations, so the 27 networks rest on nine independent geometries and the three coordinations of one geometry are nested prunings of it; each was then analyzed with two orderings of the accessible set, outermost first and random, giving 54 sweeps. The boundary stiffness matrix is the Schur complement of the stiffness matrix onto the displacement degrees of freedom of the accessible nodes; its Jacobian rows are indexed by pairs of boundary degrees of freedom, its three rigid-body zero modes are exact zero modes of every column and so need no removal before counting. The scalar control uses the same graph, the same accessible sets, and the same conductance draw. The network of Fig. 3c is one further instance of the same construction at \(7\times7\) sites, outside the \(5\times5\) to \(7\times6\) range of the ensemble; after pruning it has 47 nodes, 116 springs and mean coordination 4.94, and its electrodes are added in the radial order, farthest from the centroid first. The networks used to test the exposure bound in the elastic case are separate and are built to order: \(h\) pocket nodes on a small circle at the center, \(d\) accessible nodes on a unit circle around them, every pocket node joined to every accessible node, positions jittered by 0.02 to remove accidental collinearity, and the pocket's internal wiring empty, a path or a clique. The scalar blind dimension of the same graph is computed from the same edge list, so the comparison is on identical wiring.

\section{Hardware trajectories}\label{app:hardware}

The data are those of Ref.~\cite{dillavou2024}, deposited at Dryad~\cite{dryad2024}. The deposit also holds one XOR file with six accessible nodes and 61 steps; it is excluded from everything quoted here, because its accessible set and its task differ from the regression series. The 190 regression files are the population; 189 have a plateau, 179 a learning phase, and the 178 that have both carry every paired number quoted above. The 190 files of the regression series contain, per measurement step, the gate voltages of the 16 horizontal and 16 vertical edges of a \(4\times4\) lattice with periodic connections (NSIZE = {[}4, 4{]}, ISPERIODIC = {[}1, 1{]}) and the node voltages of the free network. The convention that the horizontal array element joins nodes \((r,c)\) and \((r,c+1)\) and the vertical element joins \((r,c)\) and \((r+1,c)\), indices modulo 4, was chosen among six candidates by the Kirchhoff current residual at the nodes that are neither driven nor read. The node voltages used are the TrainFreeState array averaged over its drive vectors. The residual of a step is the mean absolute nodal current there divided by the mean absolute branch current. The figure quoted for a convention is the median over steps and then over the first 20 files. The chosen convention gives 0.153, the next best 0.535, and the remaining four 0.74 to 1.12. Conductances were reconstructed as \(K=\beta(G-V_T-\bar V)\) with \(\beta=8\times10^{-4}\) (V\(\Omega\))\(^{-1}\), \(V_T=0.7\) V and \(\bar V\) the mean voltage of the two end nodes~\cite{dillavou2024}, floored at \(10^{-6}\). Over all 190 files the residual of the chosen convention has median 0.146, quartiles 0.131 and 0.190, and maximum 0.253, so the reconstruction carries a model error of roughly 15\%. Conductances at the floor make up 10.2\% of the values on average, from 5.1\% to 15.7\% by file, and every file has some. The accessible set is the union of the source and target nodes. For each consecutive pair of measurement steps the increment \(\Delta K\) was projected onto \(\ker J_\Lambda(B)\) evaluated at the mean of the two states, and the relative increment \(\Delta K\)/k onto the kernel of \(J_\Lambda\) diag(k) at the same state; both quantities are reported, in Table II; the isotropic expectation is the trace of the projector restricted to the edges that moved, divided by their number. The two coordinate systems agree during learning and disagree in the plateau, and both are reported. The learning-phase agreement follows from the restriction, not coincidence: the median learning step has all 32 conductances moving in 167 of 178 experiments, so the restricted trace equals the full trace, \(\dim\ker J_\Lambda\), in both coordinate systems, giving the same isotropic value (0.812 each); in the plateau typically three quarters move (median 24.75 of 32), the two coordinate systems' projectors restrict to that smaller set unequally, and the values diverge (0.849 against 0.786). The plateau was defined as the first step from which the training error stays within a factor of two of its final value. The pre-specified prediction was that the fraction of the plateau drift lying in the blind subspace exceeds the isotropic expectation. It was tested in the four combinations of subspace and plateau definition that the record allows: the subspace as pre-specified was \(\ker J_\Lambda\oplus\operatorname{span}(k)\), with \(k\) the direction of a uniform rescaling of every conductance, and the plateau as pre-specified was the second half of each run. That subspace over the second half gives 46 of 190 experiments above isotropy, and over the error-defined plateau 44 of 178. The kernel alone over the second half gives 11 of 190, and over the error-defined plateau 10 of 178. The prediction needed a majority in each case and failed in all four, and the deviation runs the other way. The four counts are in the deposited \texttt{hardware\_stats.json} under \texttt{prereg\_test}, and the prediction and its outcome are in the pre-specification record deposited with the code.

\section{Data availability}

The hardware trajectories are those of Dillavou et al.~\cite{dillavou2024}, available at Dryad (doi:10.5061/dryad.8w9ghx3vx)~\cite{dryad2024}. The edge lists and accessible sets of the exploration and held-out ensembles are deposited with the code at Zenodo (doi:10.5281/zenodo.22744869). The exploration cases carry their blind dimension. The held-out cases are deposited unlabeled, as they were generated, and the held-out evaluation is deposited beside them with the numerical and the forest blind dimension of each of the 1,500 cases scored. The networks that test the exposure bound are deposited as tables of their generating parameters with the rank and the bound of each case, and are regenerated from those parameters and the seeds in the deposited scripts; the 400 held-out networks among them are cases of the deposited held-out ensemble. The 1,839 networks of the placement survey and the 27 spring networks are deposited as their summary tables and are regenerated from the seeds in the deposited scripts, not stored as edge lists.

\section{Code availability}

Python code for the response map, the decomposition, the one-node formula, the exposure bound with its saturation and elastic tests, the tropical search, the ensembles, the placement survey, the spring networks, and the analysis of the hardware trajectories is deposited at Zenodo (doi:10.5281/zenodo.22744869), together with a record of the criteria fixed before each run and of every deviation from them, the discarded first held-out run, the script that labels the held-out ensemble, and the script that recomputes the verifications quoted in the Results.

\end{document}